\documentclass{elsart}
\usepackage{graphicx}
\usepackage{amsmath}
\usepackage{amsfonts}
\usepackage{amssymb}

\begin{document}
\begin{frontmatter}
\title{Quasi-sparse eigenvector diagonalization and stochastic error correction}
\author{Dean Lee}
\address{Dept. of Physics, Univ. of Massachusetts, Amherst, MA 01003,\\
dlee@physics.umass.edu}
\begin{abstract}
We briefly review the diagonalization of quantum Hamiltonians using the
quasi-sparse eigenvector (QSE) method. \ We also introduce the technique of
stochastic error correction, which systematically removes the truncation error
of the QSE result by stochastically sampling the contribution of the remaining
basis states.
\end{abstract}
\end{frontmatter}

\section{Quasi-sparse eigenvector method}

Quasi-sparse eigenvector (QSE) diagonalization is a new computational method
which finds approximate low-lying eigenvalues and eigenvectors for a general
quantum field Hamiltonian $H$ \cite{qse}. \ It handles the exponential
increase in the dimension of Fock space by exploiting the sparsity of the
Hamiltonian. \ The method is most effective when the splitting between
low-lying eigenvalues is not too small compared to the size of the Hamiltonian
off-diagonal entries. \ In such cases the low-lying eigenvectors are
quasi-sparse, meaning that the vector is dominated by a small fraction of its
largest components. \ The QSE algorithm is applied using the following steps:

\begin{enumerate}
\item  Select a subset of orthonormal basis vectors $\left\{  e_{i_{1}}%
,\cdots,e_{i_{n}}\right\}  $ and call the corresponding subspace $S$.

\item  Diagonalize $H$ restricted to $S$ and find one eigenvector $v.$

\item  Sort the basis components $\langle e_{i_{j}}|v\rangle$ according to
their magnitude and remove the least important basis vectors.

\item  Replace the discarded basis vectors by new basis vectors. \ These are
selected at random from a pool of candidate basis vectors which are connected
to the old basis vectors through non-vanishing matrix elements of $H$.

\item  Redefine $S$ as the subspace spanned by the updated set of basis
vectors and repeat steps 2 through 5.
\end{enumerate}

\noindent

If the subset of basis vectors is sufficiently large, the exact low energy
eigenvectors will be stable fixed points of the QSE update process. \ We can
show this as follows. \ Let $\left|  i\right\rangle $ be the eigenvectors of
the submatrix of $H$ restricted to the subspace $S$, where $S$ is the span of
the subset of basis vectors after step 3 of the QSE algorithm. \ Let $\left|
A_{j}\right\rangle $ be the remaining orthonormal basis vectors in the full
space not contained in $S$. \ We can represent $H$ as%
\begin{equation}
\left[
\begin{array}
[c]{cccccc}%
\lambda_{1} & 0 & \cdots & \left\langle 1\right|  H\left|  A_{1}\right\rangle
& \left\langle 1\right|  H\left|  A_{2}\right\rangle  & \cdots\\
0 & \lambda_{2} & \cdots & \left\langle 2\right|  H\left|  A_{1}\right\rangle
& \left\langle 2\right|  H\left|  A_{2}\right\rangle  & \cdots\\
\vdots & \vdots & \ddots & \vdots & \vdots & \cdots\\
\left\langle A_{1}\right|  H\left|  1\right\rangle  & \left\langle
A_{1}\right|  H\left|  2\right\rangle  & \cdots &  E\cdot\lambda_{A_{1}} &
\left\langle A_{1}\right|  H\left|  A_{2}\right\rangle  & \cdots\\
\left\langle A_{2}\right|  H\left|  1\right\rangle  & \left\langle
A_{2}\right|  H\left|  2\right\rangle  & \cdots & \left\langle A_{2}\right|
H\left|  A_{1}\right\rangle  & E\cdot\lambda_{A_{2}} & \cdots\\
\vdots & \vdots & \vdots & \vdots & \vdots & \ddots
\end{array}
\right]  .
\end{equation}
We have written the diagonal terms for the basis vectors $\left|
A_{j}\right\rangle $ with an explicit factor $E$ for reasons to be explained
shortly. \ We let $\left|  1\right\rangle $ be the approximate eigenvector of
interest and shift the diagonal entries so that $\lambda_{1}=0.$ \ Our
starting hypothesis is that $\left|  1\right\rangle $ is close to some exact
eigenvector of $H$ which we denote as $\left|  1_{\text{full}}\right\rangle $.
\ More precisely we assume that the components of $\left|  1_{\text{full}%
}\right\rangle $ outside $S$ are small enough so that we can expand in inverse
powers of the introduced parameter $E.$

We now expand the eigenvector as%
\begin{equation}
\left|  1_{\text{full}}\right\rangle =\left[
\begin{array}
[c]{c}%
1+\cdots\\
c_{2}^{\prime}E^{-1}+\cdots\\
\vdots\\
c_{A_{1}}^{\prime}E^{-1}+\cdots\\
c_{A_{2}}^{\prime}E^{-1}+\cdots\\
\vdots
\end{array}
\right]  \label{eigvec}%
\end{equation}
and the corresponding eigenvalue as%
\begin{equation}
\lambda_{\text{full}}=\lambda_{1}^{\prime}E^{-1}+\cdots.
\end{equation}
In (\ref{eigvec}) we have constrained $\left|  1_{\text{full}}\right\rangle $
to have unit norm. \ From the eigenvalue equation $H\left|  1_{\text{full}%
}\right\rangle =\lambda_{\text{full}}\left|  1_{\text{full}}\right\rangle $ we
find at lowest order%
\begin{equation}
c_{A_{j}}^{\prime}=-\tfrac{\left\langle A_{j}\right|  H\left|  1\right\rangle
}{\lambda_{A_{j}}}. \label{z}%
\end{equation}
We see that at lowest order the component of $\left|  1_{\text{full}%
}\right\rangle $ in the $\left|  A_{j}\right\rangle $ direction is independent
of the other vectors $\left|  A_{j^{\prime}}\right\rangle $. \ If $\left|
1\right\rangle $ is sufficiently close to $\left|  1_{\text{full}%
}\right\rangle $ then the limitation that only a fixed number of new basis
vectors is added in step 4 of the QSE algorithm is not important. \ At lowest
order in $E^{-1}$ the comparison of basis components in step 3 (in the next
iteration) is the same as if we had included all remaining vectors $\left|
A_{j}\right\rangle $ at once. \ Therefore at each update only the truly
largest components are kept and the algorithm will converge to some optimal
approximation of $\left|  1_{\text{full}}\right\rangle $. \ This is consistent
with the actual performance of the algorithm for a wide range of examples.

\section{Stochastic error correction}

The quasi-sparse eigenvector method approximates the low energy eigenvalues
and eigenvectors of a Hamiltonian $H$ by finding and diagonalizing optimized
subspaces $S$. \ This procedure however has an intrinsic error due to the
omission of basis states not in $S$. \ The goal of stochastic error correction
is to sample the contribution of these remaining basis vectors in a systematic
fashion. \ Due to space limitations we consider here only the series expansion
method. \ Several other techniques are discussed in a forthcoming paper
\cite{sec}.

The series method for stochastic error correction is based on the $E^{-1}$
expansion introduced in the previous section,
\begin{equation}
\left|  1_{\text{full}}\right\rangle \propto\left[
\begin{array}
[c]{c}%
1\\
c_{2}^{\prime}E^{-1}+c_{2}^{\prime\prime}E^{-2}+\cdots\\
\vdots\\
c_{A_{1}}^{\prime}E^{-1}+c_{A_{1}}^{\prime\prime}E^{-2}+\cdots\\
c_{A_{2}}^{\prime}E^{-1}+c_{A_{2}}^{\prime\prime}E^{-2}+\cdots\\
\vdots
\end{array}
\right]  , \label{unnorm}%
\end{equation}%
\begin{equation}
\lambda_{\text{full}}=\lambda_{1}^{\prime}E^{-1}+\lambda_{1}^{\prime\prime
}E^{-2}\cdots.
\end{equation}
This time it is more convenient to choose the normalization of the eigenvector
such that the $\left|  1\right\rangle $ component remains 1. \ The convergence
of the expansion is controlled by the proximity of $\left|  1\right\rangle $
to $\left|  1_{\text{full}}\right\rangle $. \ If $\left|  1\right\rangle $ is
not at all close $\left|  1_{\text{full}}\right\rangle $ then it will be
necessary to use a more robust method, and such methods are discussed in
\cite{sec}.

At first order in $E^{-1}$ we find%
\begin{equation}
c_{A_{j}}^{\prime}=-\tfrac{\left\langle A_{j}\right|  H\left|  1\right\rangle
}{\lambda_{A_{j}}}%
\end{equation}%
\begin{equation}
\lambda_{1}^{\prime}=-\sum_{j}\tfrac{\left\langle 1\right|  H\left|
A_{j}\right\rangle \left\langle A_{j}\right|  H\left|  1\right\rangle
}{\lambda_{A_{j}}}%
\end{equation}%
\begin{equation}
c_{j}^{\prime}=\tfrac{1}{\lambda_{j}}\sum_{k}\tfrac{\left\langle j\right|
H\left|  A_{k}\right\rangle \left\langle A_{k}\right|  H\left|  1\right\rangle
}{\lambda_{A_{k}}}.
\end{equation}
The second order contributions are%
\begin{equation}
c_{A_{j}}^{\prime\prime}=\sum_{k\neq j}\tfrac{\left\langle A_{j}\right|
H\left|  A_{k}\right\rangle \left\langle A_{k}\right|  H\left|  1\right\rangle
}{\lambda_{A_{j}}\lambda_{A_{k}}}-\sum_{l\neq1}\sum_{k}\tfrac{\left\langle
A_{j}\right|  H\left|  l\right\rangle \left\langle l\right|  H\left|
A_{k}\right\rangle \left\langle A_{k}\right|  H\left|  1\right\rangle
}{\lambda_{A_{j}}\lambda_{l}\lambda_{A_{k}}}%
\end{equation}
\qquad%
\begin{equation}
\lambda_{1}^{\prime\prime}=\sum_{j}\sum_{k\neq j}\tfrac{\left\langle 1\right|
H\left|  A_{j}\right\rangle \left\langle A_{j}\right|  H\left|  A_{k}%
\right\rangle \left\langle A_{k}\right|  H\left|  1\right\rangle }%
{\lambda_{A_{j}}\lambda_{A_{k}}}-\sum_{j}\sum_{l\neq1}\sum_{k}\tfrac
{\left\langle 1\right|  H\left|  A_{j}\right\rangle \left\langle A_{j}\right|
H\left|  l\right\rangle \left\langle l\right|  H\left|  A_{k}\right\rangle
\left\langle A_{k}\right|  H\left|  1\right\rangle }{\lambda_{A_{j}}%
\lambda_{l}\lambda_{A_{k}}}%
\end{equation}%

\begin{align}
c_{m}^{\prime\prime}  &  =-\sum_{j}\sum_{k\neq j}\tfrac{\left\langle m\right|
H\left|  A_{j}\right\rangle \left\langle A_{j}\right|  H\left|  A_{k}%
\right\rangle \left\langle A_{k}\right|  H\left|  1\right\rangle }{\lambda
_{m}\lambda_{A_{j}}\lambda_{A_{k}}}\\
&  +\sum_{j}\sum_{l\neq1}\sum_{k}\tfrac{\left\langle m\right|  H\left|
A_{j}\right\rangle \left\langle A_{j}\right|  H\left|  l\right\rangle
\left\langle l\right|  H\left|  A_{k}\right\rangle \left\langle A_{k}\right|
H\left|  1\right\rangle }{\lambda_{m}\lambda_{A_{j}}\lambda_{l}\lambda_{A_{k}%
}}\nonumber\\
&  -\tfrac{1}{\lambda_{m}}\left[  \sum_{j}\tfrac{\left\langle 1\right|
H\left|  A_{j}\right\rangle \left\langle A_{j}\right|  H\left|  1\right\rangle
}{\lambda_{A_{j}}}\right]  \tfrac{1}{\lambda_{m}}\left[  \sum_{k}%
\tfrac{\left\langle m\right|  H\left|  A_{k}\right\rangle \left\langle
A_{k}\right|  H\left|  1\right\rangle }{\lambda_{A_{k}}}\right]  .\nonumber
\end{align}

These contributions are calculated by a Monte Carlo sampling process. \ Let
$P(A_{\text{trial}(i)})$ denote the probability of selecting $A_{\text{trial}%
(i)}$ on the $i^{\text{th}}$ trial. \ For example the first order correction
to the eigenvalue can be calculated using
\begin{align}
\lambda_{1}^{\prime} &  =-\sum_{j}\tfrac{\left\langle 1\right|  H\left|
A_{j}\right\rangle \left\langle A_{j}\right|  H\left|  1\right\rangle
}{\lambda_{A_{j}}}\\
&  =-\lim_{N\rightarrow\infty}\tfrac{1}{N}\sum_{i=1,\cdots,N}\tfrac
{\left\langle 1\right|  H\left|  A_{\text{trial}(i)}\right\rangle \left\langle
A_{\text{trial}(i)}\right|  H\left|  1\right\rangle }{\lambda_{A_{\text{trial}%
(i)}}P(A_{\text{trial}(i)})}.
\end{align}
Application of the series method for stochastic error correction to the
Hubbard model was discussed by Nathan Salwen at the Heidelberg meeting and is
presented in another article in these proceedings.

\texttt{ }

\paragraph*{Acknowledgments}

The author thanks the organizers of the Light-Cone Meeting at Heidelberg and
collaborators on the original papers cited here. \ Financial support provided
by the National Science Foundation.

\end{document}